# Merohedral disorder and impurity impacts on superconductivity of fullerenes

Shu-Ze Wang[1*], Ming-Qiang Ren[1*], Sha Han[1], Fang-Jun Cheng[1], Xu-Cun Ma[1,2], Qi-Kun Xue[1,2,3,4], Can-Li Song[1,2†]

[1]*State Key Laboratory of Low-Dimensional Quantum Physics, Department of Physics, Tsinghua University, Beijing 100084, China*

[2]*Frontier Science Center for Quantum Information, Beijing 100084, China*

[3]*Beijing Academy of Quantum Information Sciences, Beijing 100193, China*

[4]*Southern University of Science and Technology, Shenzhen 518055, China*

Local quasiparticle states around impurities provide essential insight into the mechanism of unconventional superconductivity, especially when the candidate materials are proximate to an antiferromagnetic Mott-insulating phase. While such states have been reported in atom-based cuprates and iron-based compounds, they are unexplored in organic superconductors which feature tunable molecular orientation. Here we employ scanning tunneling microscopy and spectroscopy to reveal multiple forms of robustness of an exotic *s*-wave superconductivity in epitaxial $Rb_3C_{60}$ films against merohedral disorder, non-magnetic single impurities and step edges at the atomic scale. Also observed have been Yu-Shiba-Rusinov (YSR) states induced by deliberately incurred Fe adatoms that act as magnetic scatterers. The bound states display abrupt spatial decay and vary in energy with the Fe adatom registry. Our results and the universal optimal superconductivity at half-filling point towards local electron pairing in which the multiorbital electronic correlations and intramolecular phonons together drive the high-temperature superconductivity of doped fullerenes.

*Both authors contributed equally to this work.

†*To whom correspondence should be addressed. Email: clsong07@mail.tsinghua.edu.cn*

Disorder, impurities in an otherwise homogeneous superconductor, are often undesired aliens because they may hinder observations of intrinsic properties of the host material[1-3]. Yet dopant impurities could also be a double-edged sword by leading not only to emergent high-temperature ($T_c$) superconductivity in cuprates and iron pnictides[4,5] but also to uncovering the underlying mechanism of unconventional superconductivity[6-10], especially as multiple unusual states are complexly intertwined in these correlated electron materials[11,12]. Whereas it has been well documented that non-magnetic impurities little affect Cooper pairs in conventional superconductors[13,14], they turn out to induce local bound states inside the superconducting gap and suppress superconductivity via pair breaking for unconventional pairing symmetries, for example, in a *d*-wave or $s_\pm$ wave superconductor[7,14-17]. Recently, anomalous enhancement of superconductivity by disorder is another example of impurities revealing their fundamental significance for the low-dimensional superconducutors[18-20]. It is therefore tempting to consider impurities as a blessing in disguise to understand the physics of candidate superconductors[4-17], to strive for optimal superconductivity[21], and to create electronic states that never emerge from pure superconducting systems[22].

Unlike atom-based superconductors, an organic superconductor is a synthetic molecule-based compound that uniquely exhibits additional degrees of freedom related to its molecular orientation. Consequently, inequivalent molecular orientations might unavoidably take place. Such orientational (merohedral) disorder has been seen early in pure and doped fullerenes[23,24], but its impact on superconductivity of fullerenes is highly controversial[25-28]. Furthermore, the fullerides represent an unusual category of organic superconductors in which the multiorbital electronic correlations and electron-phonon interactions are both suggested to be essentially significant to reach high-$T_c$ superconductivity[29,30]. In this context, a systematic investigation of impurity effects on superconductivity of doped fullerenes would provide justification of the previously advocated *s*-wave pairing symmetry[31-33], as well as advance the understanding of their exotic superconducting state. However, such experiment was unexplored and the roles played by magnetic and non-magnetic impurities remain lacking in fulleride superconductors.

In this work, we use a state-of-the-art molecular beam epitaxy (MBE) technique to grow epitaxial films of rubidium(Rb) doped fullerenes with thickness and filling tunability, and probe the local quasiparticle states in the vicinity of various impurities at the atomic scale by means of cryogenic scanning tunneling microscopy (STM) and spectroscopy (STS). Distinct from both the superconducting $K_3C_{60}$ films without merohedral disorder[33] and insulating $Cs_3C_{60}$ ones with great merohedral disorder[34], the merohedrally disordered $Rb_3C_{60}$ films are superconducting. This allows for atomic-scale visualization of merohedral disorder impact on superconductivity of fullerenes, which, together with a detailed STS investigation on magnetic and non-magnetic impurities, shows that superconductivity of fullerenes is entirely consistent with local *s*-wave pairing. By exploring the thickness and electron filling variations of superconducting gap $\Delta$, we establish a unified phase diagram of fullerenes in which the optimal superconductivity always develops at half-filling.

**Results**

**Merohedral disorder and its impact on superconductivity**

Figure 1a depicts a representative STM topography of 9-monolayer (ML) $Rb_3C_{60}$ thin films epitaxially grown on graphitized SiC(0001) substrates. Evidently, not all $C_{60}$ molecules have the same orientation, although one three-fold symmetry axis for every $C_{60}$ is always perpendicular to the surface. More specifically, many nanoscale domains with two distinct $C_{60}$ orientations, related by 44.48° rotation about the [111] axis (Fig. 1b), develop and are partially opacified in red and orange, respectively. This is reminiscent of the two standard orientations of $C_{60}$ that occur randomly and cause great merohedral disorder in face-centered cubic (fcc)-structured trivalent fullerides[24-26]. The $C_{60}$ orientations are more disordered in regions between adjacent merohedral domains. In order to quantify the merohedral disorder, we calculate the averaged orientational correlation function $<\cos(\theta_{ij})>$[35], in which $\theta_{ij} = \theta_i - \theta_j$ denotes the angle between nearest neighbor $C_{60}$ molecules (i.e. i and j), and summarize them in Fig. 1c. As the alkali metals increase in atomic radius, a decrease of orientational correlation means increased merohedral disorder. This is primarily caused by the weaking of Coulomb repulsions between neighboring trivalent $C_{60}$ ions associated with a lattice expansion[36], which otherwise stabilize a long-ranged merohedral order in the $K_3C_{60}$ films[33], to wit $<\cos(\theta_{ij})> = 1$.

Tunneling spectroscopy of fullerides probes the local density of quasiparticle states (DOS) and measures the superconducting energy gap at the Fermi level ($E_F$). In Fig. 1d, we compare the tunneling $dI/dV$ spectra on 9 ML trivalent fulleride films $A_3C_{60}$ ($A$ = K, Rb, Cs) doped with different alkali metals. A noticeable observation is that the merohedral disorder blurs the two sharp DOS peaks around -0.4 eV and 0.1 eV, which happen exclusively for the merohedrally ordered $K_3C_{60}$ films (top curve)[33]. This results in a generally smooth variation of electronic DOS in both $Rb_3C_{60}$ and $RbCs_2C_{60}$ films, as theoretically anticipated[37]. It is, however, worth noting that the low-lying DOS width roughly estimated as the spacing between the two conductance minima below and above $E_F$ (> 1.2 eV, see the two dashed lines in Fig. 1d) is significantly larger than the commonly argued $t_{1u}$ bandwidth of ~ 0.5 eV[37,38]. Such a discrepancy should originate from the Jahn-Teller (JT) instabilities and Coulomb interactions omitted by three-band first-principles calculations[37,38]. In consideration of JT-induced subband splitting and electronic correlations, the $t_{1u}$ bandwidth would be substantially increased[30,39,40] and accords with our observations. A further enhancement of electronic correlations $U$ pushes the $t_{1u}$-derived DOS toward higher energy and concurrently opens a Mott insulating gap in the most expanded $Cs_3C_{60}$ films (see the bottom curve in Fig. 1d)[34].

Despite significant variation in low-lying electronic DOS, superconductivity develops well in $Rb_3C_{60}$ films and has little to do with the nanoscale merohedral disorder. We unambiguously reveal this by measuring the spatial dependence of superconducting gaps at the atomic scale via STS, as exemplified in Fig. 1e. Even on the regions between nearest neighbor merohedral domains, the superconducting gaps exhibit apparent coherence peaks (blue curves) and are immune to the local merohedral disorder (Fig. 1a). This is more compellingly confirmed in Cs and Rb co-doped $RbCs_2C_{60}$ films, which imprint a somewhat stronger merohedral disorder but still exhibit a superconducting transition temperature up to $T_c$ = 23 K (Supplementary Fig. 1). Nevertheless, the d$I$/d$V$ spectra present some spatial electronic inhomogeneities, especially in the superconducting coherence peak. A careful examination of the superconducting $Rb_3C_{60}$ films at varied thicknesses and spatial locations intriguingly reveals that the coherence peak amplitude scales inversely with the gap size $\Delta$ (Supplementary Fig. 2). This is unexpected by the conventional wisdom of Bardeen-Cooper-Schrieffer (BCS) picture, and mostly ascribed to a coexistence of competing order, e.g. the ubiquitous pseudogap phase (Supplementary Fig. 3)[33]. Similar behavior and pseudogap phenomenology have been well documented in cuprate superconductors[41].

**Thickness and filling dependence of superconductivity**

Having established the merohedral disorder-independent superconductivity in trivalent fullerides, we then explore its dependence on thickness and electron filling. Plotted in Fig. 2a-c is the temperature variation of spatially-averaged tunneling spectra measured on 3 ML, 6 ML and 9 ML $Rb_3C_{60}$, respectively. Again, the fully gapped superconductivity with an isotropic *s*-wave gap function is consistently demonstrated in the superconducting $Rb_3C_{60}$ films and gets smeared out at elevated temperatures. By examining the temperature dependence of the gap depth in Fig. 2d, the critical temperature $T_c$ is determined and increases from 23 K for 3 ML, to 26 K for 6 ML, and 28 K for 9 ML $Rb_3C_{60}$. Such a $T_c$ evolution with film thickness stands in marked contrast to $K_3C_{60}$, where the maximum $T_c$ occurs in 3 ML films[33]. This hints at another factor, possibly linking with the alkali metal ion size-dependent electronic correlations ($U$) of $A_3C_{60}$[32], to consider for a complete understanding of superconductivity of $A_3C_{60}$ films at varied thicknesses. Note that monolayer and bilayer $Rb_3C_{60}$ films are non-superconducting at all due to the significantly increased electronic correlations (Supplementary Fig. 4), in analogy to $K_3C_{60}$ counterparts[33]. A residual DOS depletion around $E_F$, hence the pseudogap, is also observed in superconducting $Rb_3C_{60}$ thin films above $T_c$ (see the red curves in Fig. 2a-c) and within vortices (Supplementary Fig. 3b).

In what follows, we explore the superconductivity of $Rb_xC_{60}$ by tuning the stoichiometry and thus electron filling *x*. Figure 2e summarizes the superconducting gap Δ (top panel) and averaged orientational correlation (bottom panel) as a function of Rb doping level *x*. Clearly, Δ increases with the film thickness, in good accordance with $T_c$ (Fig. 2d). The extracted reduced gap ratio $2\Delta/k_BT_c = 6.0 \pm 0.4$ is comparable to that of $K_xC_{60}$ films[33], but appreciably exceeds the canonical BCS value of 3.53. More interestingly, $\Delta$ invariably reaches its peak at half-filling irrespective of film thickness, and declines more quickly below half-filling for thin $Rb_3C_{60}$ films. Notwithstanding a dome-shaped variation of $\Delta$, the merohedral disorder remains essentially unchanged with electron filling $x$ and film thickness (bottom panel of Fig. 1e). This not only corroborates the above claim that superconductivity is unaffected by merohedral disorder[27,42], but also hints that the dome-shaped superconducting phase diagram does not correlate from any $x$-dependent merohedral disorder effects.

**Robust superconductivity against non-magnetic impurities**

As the electron doping of $Rb_xC_{60}$ deviates slightly from half-filling, subsurface tetragonal Rb vacancies emerge as dark windmills as $x < 3$, whereas excess K adatoms appear and occupy the octahedral sites as $x > 3$. Analogous to $K_xC_{60}$ (Supplementary Fig. 5a,b)[33], they do not alter profoundly the orientation of the nearby fullerene molecules and thus serve as intrinsically non-magnetic impurities to test the fully gapped superconductivity in fullerides. Figure 3a-d show the STM topographies of a single Rb vacancy and an excess Rb adatom, as well as linecut d$I$/d$V$ spectra taken across both impurities. No evidence of in-gap bound state is revealed (red curves), although the gap size $\Delta$ shrinks by $\sim$ 25% on Rb excess impurity. Similar responses of the superconducting gap to K impurities have been observed in $K_xC_{60}$ as well (Supplementary Fig. 5c). Here the $\Delta$ reduction probably arises from a local doping variation, namely a deviation of $x$ from 3. The alkali metal vacancies are located below the top $C_{60}$ molecules, rendering the local $\Delta$ reduction invisible in surface-sensitive STS.

On the other hand, step edges could be seen as one-dimensional perturbations and bring about Andreev bound states as they are normal to the possible sign-changing direction in $\Delta$[43].

The spectroscopic signature of these bound states, e.g. a zero-bias conductance peak (ZBCP), has been observed experimentally in a few cuprate and iron-pniticide superconductors[41,44,45]. Figure 3e depicts a topographic STM image of one monomolecular step edge separating $Rb_3C_{60}$ epitaxial films between 8 ML (lower terrace) and 9 ML (upper terrace). We note that the step edge runs nearly along the close-packed directions of $C_{60}$ molecules. Figure 3f shows a series of d$I$/d$V$ spectra taken along a trajectory approaching the step edge (solid line in Fig. 3e). The superconducting gap proves to be undisturbed at the step edge and nearby, and no evidence of Andreev bound states is found. Some random variations in the coherence peak, including the pronounced coherence peaks near Rb excess impurity in Fig. 3d, might be related to the slight electronic inhomogeneity of superconducting $Rb_3C_{60}$ films (Fig. 1e).

**Local probe of Yu-Shiba-Rusinov states**

To fully understand the impurity impact on fulleride superconductivity, we intentionally deposited Fe atoms on $Rb_3C_{60}$ surface at low temperature (~ 100 K). Single Fe adatoms formed (bright protrusions) and occupied top or hollow sites of the surface $C_{60}$ lattice, dubbed as Fe(I) and Fe(II) in Fig. 4a. Figure 4b represents the d$I$/d$V$ spectra on both Fe impurities and defect-free regions. Note that multiple Fe(I) and Fe(II) impurities have been measured and averaged to eliminate the spatial inhomogeneity effects of d$I$/d$V$ spectra. Evidently, both Fe(I) and Fe(II) adatoms act as magnetic scatterers and significantly suppress the superconducting coherence peaks, while a prominent zero-bias conductance peak (ZBCP) develops on Fe(I). They are the hallmarks of Yu-Shiba-Rusinov (YSR) states induced by a coupling of magnetic impurity to an $s$-wave superconductor[8,10,13,14,22,46-49]. Figure 4c shows a series of tunneling spectra across an isolated Fe(I) impurity. The ZBCP intensity decreases quite abruptly and becomes barely visible at a spatial distance of 1.4 nm from the impurity site. Here the distinct behaviors of YSR states on Fe(I) and Fe(II) may be caused by the varied coupling strength between them and the Rb$_3$C$_{60}$ films[50]. In other words, the exchange coupling of Fe(II) adsorbed at the hollow sites to Copper pairs might be so significantly weak that the YSR states nearly merge into the superconducting gap edges and are little discernible. Further theoretical analysis is needed to comprehensively understand the Fe registry site-dependent YSR bound states in fulleride superconductors.

**Discussion**

Our atomic-scale observations of short-range YSR bound states solely on single magnetic Fe adatoms, in conjunction with the multiple forms of robustness of superconductivity against non-magnetic merohedral disorder and impurities, unambiguously confirm a sign-unchanged *s*-wave pairing state in fulleride superconductors. Distinct from a conventional superconductor, however, in charged fullerenes the $t_{1u}$-derived conduction band of ~ 0.5 eV is narrow and comparable to the electron-vibron interactions, thereby causing a breakdown of the Migdal's theorem[51-53]. As a result, superconductivity with local nonretarded attractive interactions[29,30] would be less sensitive to distribution of the electronic DOS in conduction band[54], and instead determined by some ensemble averaged DOS[55]. This differs from the usual BCS superconductor where $T_c$ is essentially governed by the DOS at $E_F$, and happens to match the non-correlated superconductivity with the merohedral disorder that significantly change the $t_{1u}$-derived DOS distribution (Fig. 1d). Moreover, such local electron pairing[29,30], mediated by intramolecular JT phonons[56,57], has also been reinforced by a short coherence length. As estimated from the vortex core radius (Supplementary Fig. 3a,c), the coherence length of a Cooper pair is 1.5 ± 0.2 nm in Rb$_3$C$_{60}$ and 2.6 ± 0.5 nm in K$_3$C$_{60}$[33], respectively, which amount to only about twice the separation between nearest neighbor fullerene molecules.

In the local pairing mechanism, the key ingredients for high-$T_c$ superconductivity are the strong coupling of the $t_{1u}$ electrons to intramolecular JT phonons in trivalent fullerides[55-57]. The phonon-mediated unusual multiorbital (attractive) interactions lead to an effectively inverted Hund's coupling (S = 1/2)[32] and a local spin-singlet s-wave pairing on the same orbital[30], further enhanced via a coherent tunneling of pairs between orbitals (the Suhl-Kondo mechanism)[58,59]. On the other hand, the multiorbital electronic correlations suppress electron hopping-induced charge fluctuations and more effectively bind electrons into intraorbital pairs[29]. In this sense, the Coulomb interactions actually helps the local pairing, until they are strong enough to drive a transition from a superconductivity to Mott-insulating phase[33]. Such a local pairing scenario naturally accounts for the dome-shaped dependence of $T_c$ on $C_{60}$ packing density-controlled $U$[27,32,42] as well as the conflicting variation of superconductivity with film thickness in $K_3C_{60}$ and $Rb_3C_{60}$. In $K_3C_{60}$, $U$ is relatively small and its enhancement at reduced film thicknesses stabilizes the local pairing and thus enhances superconductivity[33], whereas the opposite holds true due to the already strong electronic correlations in $Rb_3C_{60}$. A further enhancement of $U$ pushes thin $Rb_3C_{60}$ films closer to a Mott transitions and thus weakens superconductivity, as observed above. In Fig. 5, we schematically illustrate a unified phase diagram of the charged fullerides and discover a universal optimal superconductivity at half-filling, no matter how the electronic correlations $U$ varies with the alkali metal and film thickness. This finding is unusual and most probably stems from a decrease in the dynamical JT-related pair binding energy ($U_x$, negative) away from half-filling[57,58], until the superconductivity vanishes as the $U_x$ changes its sign. For the evenly charged fullerenes, $U_2$ and $U_4$ are positive and the JT coupling instead stabilizes two correlated insulating ground states[57,58]. It is also important to note that an asymmetry of $\Delta$ versus $x$ phase diagram relative to half-filling ($x$ = 3) occurs as the electronic correlations are strong. This is related to a monotone shrinkage of $U$ with the doping $x$ in view of the enhanced Coulomb screening from itinerant electron carriers[33]. In strongly correlated regimes, a small increase of $U$ below half-filling would suppress superconductivity significantly and leads to the observed dome asymmetry.

Finally, we note that the large ratio of energy gap $\Delta$ to critical temperature (e.g. $2\Delta/k_B T_c$ > 6.0) seems to be a generic trait of high-$T_c$ superconductivity in narrow-band systems[8,17,31,33,45], including the copper-oxide superconductors[41]. Such a large deviation from the canonical BCS value of 3.53 could be straightforward and easy to understand theoretically in the framework of superconductivity with local nonretarded attractive interactions[54]. Experimentally, a local pairing mechanism, assisted cooperatively by a dynamic interfacial polaron, has been recently proposed to be responsible for the high-$T_c$ superconductivity in monolayer FeSe epitaxial films on SrTiO$_3$ substrate[60]. A question naturally arises as to whether the local paring mechanism is applicable to other narrow-band cuprates and multi-band iron pnictides[61]. Another interesting issue is to unravel the nature of pseudogap phenomenology that ubiquitously emerges from doped fullerene films (Fig. 2a-c). Whether the pseudogap shares the same mechanism as that of cupartes and how it interplays with cuprate superconductivity remain unsolved issues that merit further investigations. In any case, our experimental results of fulleride superconductors shed important light on the electron pairing in narrow-band high-$T_c$ superconductors.

**Methods**

**Sample preparations.** Our experiments were conducted in a commercial Unisoku 1500 ultra-high vacuum STM facility, connected to an MBE chamber for *in-situ* film preparation. The base pressure of both chambers is lower than $2.0 \times 10^{-10}$ Torr. $C_{60}$ molecules were evaporated from a standard Knudsen diffusion cell and grew layer-by-layer on nitrogen-doped SiC(0001) wafers (0.1 Ω·cm) at 473 K, which were pre-graphitized by thermal heating (up to 1600 K) to form a bilayer graphene. Desired alkali metal atoms (Rb or Cs) were then deposited from thoroughly outgassed SAES getters on $C_{60}$ films at a low temperature of ∼ 200 K step by step, followed by > 3 hours of post-annealing at room temperature. The layer index $n$ of $Rb_xC_{60}$ multilayers ($n \leq 3$) is determined from the step height, STM topographies and tunneling d$I$/d$V$ spectra, which varies significantly with $n$. The flux rate of $C_{60}$ is therefore calculated by dividing the coverage of fulleride films by growth time, and is used as a reference for determination of the nominal thickness for thicker fullerides, e.g. $n$ = 6 and 9 in the main text.

Away from half-filling, the electron doping $x$ is calculated directly from the areal density of Rb vacancies (Fig. 3a) or excess Rb dopants (Fig. 3c). For $x \sim 3$, there exist little defect that leads to trivalent fulleride films (Fig. 1a). A single Rb vacancy (excess) is reasonably considered as one missing (additional) Rb dopant relative to $Rb_3C_{60}$. For higher doping, $x$ is estimated from the coverage of Rb clusters, since the excess Rb dopants are individually undistinguishable. By this method, the estimated $x$ has a statistical error of < 0.5%.

**STM measurements.** After the sample growth, the fulleride epitaxial films were immediately transferred into our STM chamber for all STM and STS data collections at 4.6 K. A bias voltage was applied to the samples. To accurately characterize the superconductivity and electronic structure of fulleride films, special measures such as grounding and shielding were taken to optimize the stability and spectroscopic resolution of our STM facility. Polycrystalline PtIr tips were used after careful calibration on Ag films grown on Si(111). All STM topographic images were taken in a constant current mode. Tunneling d$I$/d$V$ spectra and electronic DOS maps were acquired using a standard lock-in technique with modulation frequency $f$ = 975 Hz, while the modulation amplitudes were 0.2 meV and 20 meV for measuring the superconducting gaps and wider-energy-range ($\pm$ 1.0 eV) d$I$/d$V$ spectra, respectively.

**Acknowledgments**

We thank H. Y. and H. W. L. for fruitful discussions. The work was financially supported by the Ministry of Science and Technology of China (2017YFA0304600, 2016YFA0301004, 2018YFA0305603), the Natural Science Foundation of China (Grants No. 51788104, No. 11634007, and No. 11774192), and in part by the Beijing Advanced Innovation Center for Future Chip (ICFC).

**Author contributions**

C.L.S., X.C.M. and Q.K.X. conceived and designed the experiments. S.Z.W., S.H. and M.Q.R. carried out the MBE growth and STM measurements. M.Q.R., S.Z.W. and C.L.S. analyzed the data and wrote the manuscript with comments from all authors.

**Additional information**

Supplementary information includes 5 figures and the corresponding figure captions.

Correspondence and requests for materials should be addressed to C.L.S.

**Competing financial interests**

The authors declare no competing financial interests.

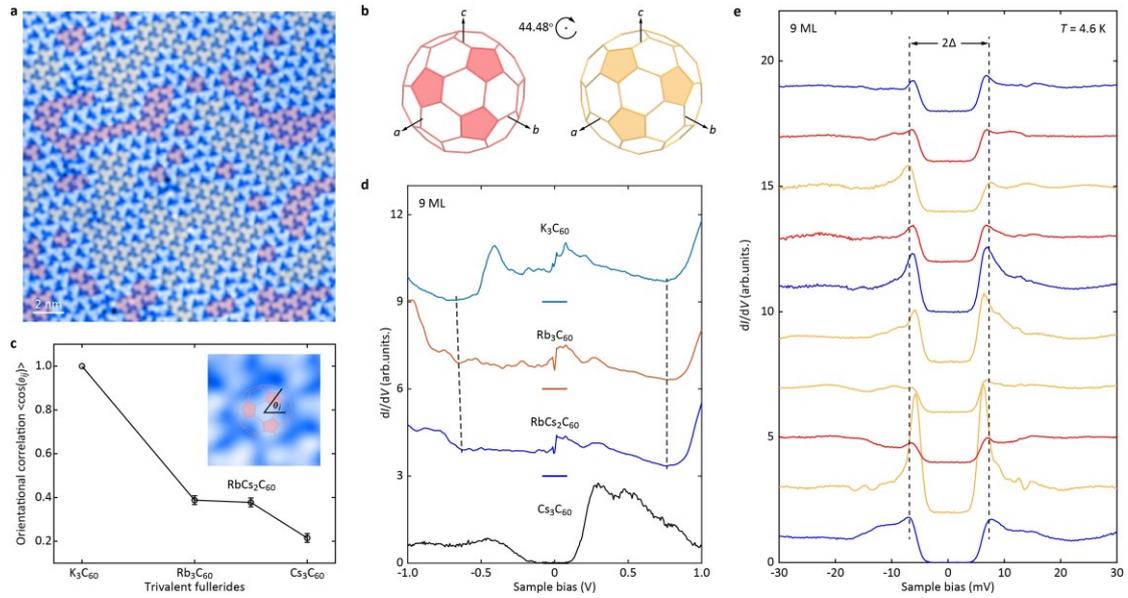

**Figure 1 Merohedral disorder and superconductivity in Rb$_3$C$_{60}$. a**, STM topography (20 nm × 20 nm, $V$ = 1.0 V, $I$ = 30 pA) on 9 ML Rb$_3$C$_{60}$ films. Two standard orientations of C$_{60}$ related by 44.48° rotation about the [111] axis are partially opacified in red and orange, respectively. **b**, Schematic view of the two standard orientations of C$_{60}$ molecules along the [111] direction. **c**, Alkali metal ion dependence of the orientational correlation of C$_{60}$ molecules (i.e. merohedral disorder). The statistical errors indicate the standard derivations of the merohedral disorder measured in various regions. Inset: definition of the angle $\theta_i$ of C$_{60}$ orientation with respect to the horizontal axis. **d**, Spatially-averaged d$I$/d$V$ spectra on 9 ML $A_3$C$_{60}$ ($A$ =K, Rb, Cs) films. The spectra have been vertically offset for clarity, with their zero conductance positions marked by correspondingly colored horizontal lines. Set point: $V$ = 1.0 V and $I$ = 200 pA. **e**, dI/dV spectra ($V$ = 30 mV and $I$ = 200 pA) taken at equal separation (2.8 nm) along the diagonal line from top left to bottom right and color coded to match the C$_{60}$ domains in **a**.

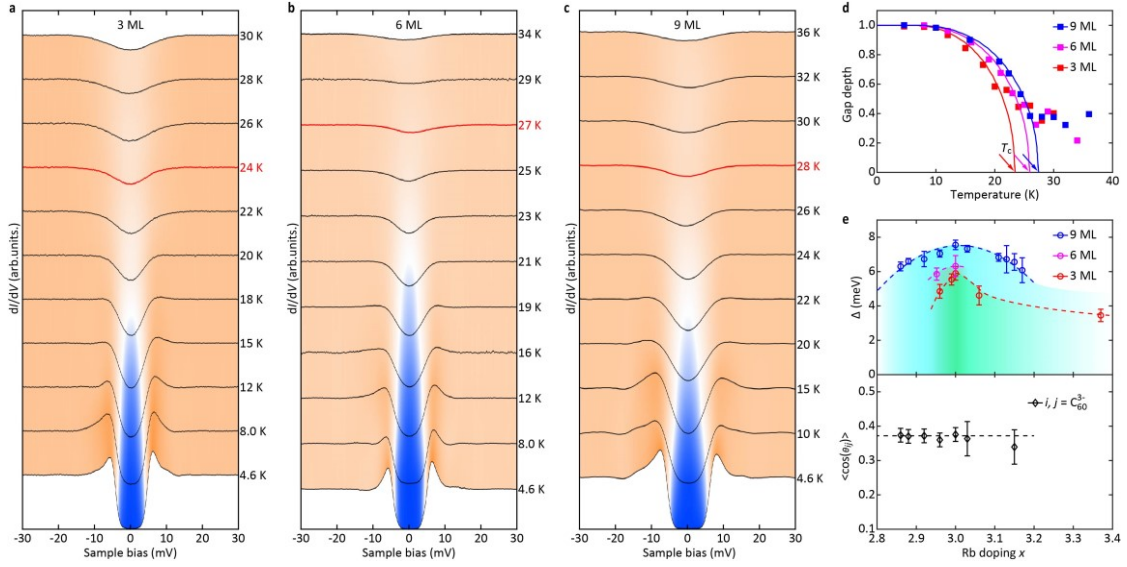

**Figure 2 Thickness dependence of superconductivity. a-c**, Spatially-averaged and normalized d$I$/d$V$ spectra as a function of temperature and thickness of Rb$_3$C$_{60}$ films as indicated. The normalization was performed by dividing the raw tunneling spectrum by its background, which was extracted from a cubic fit to the conductance for |V| > 10 mV. Set point: $V$ = 30 mV and $I$ = 200 pA. The red curves denote the residual pseudogap justly above $T_c$. **d**, Dependence of the gap depth on temperature, yielding a gradual increase of $T_c$ with film thickness (see the solid lines and arrows). **e**, Electronic phase diagram showing the evolution of $\Delta$ (empty circles) and averaged orientational correlation (diamonds) as a function of Rb doping $x$. Note that the orientational correlation function <cos($\theta_{ij}$)> is only averaged over the trivalent C$_{60}$ molecules to minimize any disruption by Rb impurities and excess atoms away from half-filling.

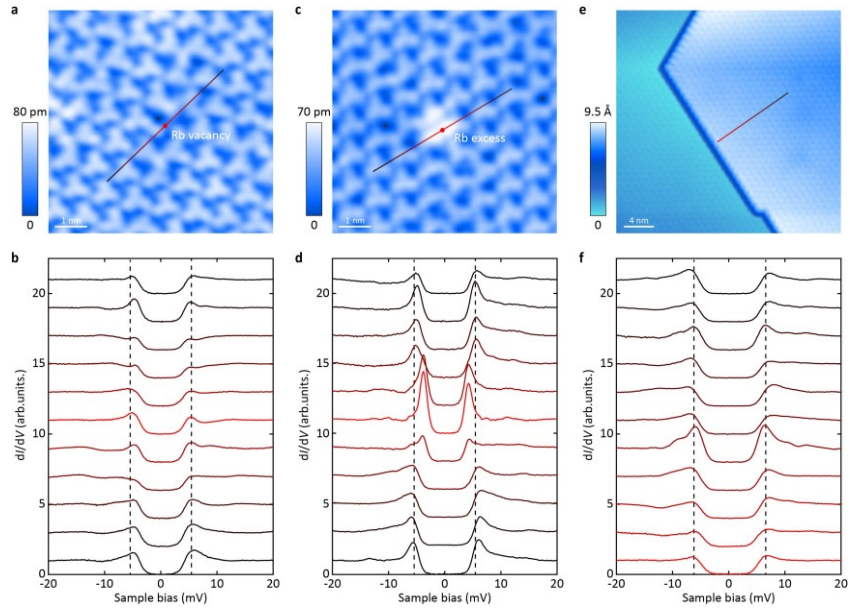

**Figure 3 Immunity of fulleride superconductivity to non-magnetic impurities. a,b**, High-resolution STM topography (7 nm × 7 nm, $V$ = 1.0 V, $I$ = 20 pA) of a single Rb vacancy (red dot) and tunneling spectra acquired 0.5 nm apart along the colored line in **a**. **c,d**, STM topography (7 nm × 7 nm, $V$ = 1.0 V, $I$ = 20 pA) of a single Rb excess impurity (red dot) and tunneling spectra acquired 0.5 nm apart along the colored line in **c**. **e,f**, STM topography (30 nm × 30 nm, $V$ = 1.0 V, $I$ = 20 pA) of $Rb_3C_{60}$ films with a monomolecular step edge and tunneling spectra acquired 1.3 nm apart along the colored line in **e**. The d$I$/d$V$ spectra are color coded to match the probed positions near (red) and away from impurities (black). Tunneling gap is set at $V$ = 30 mV and $I$ = 200 pA.

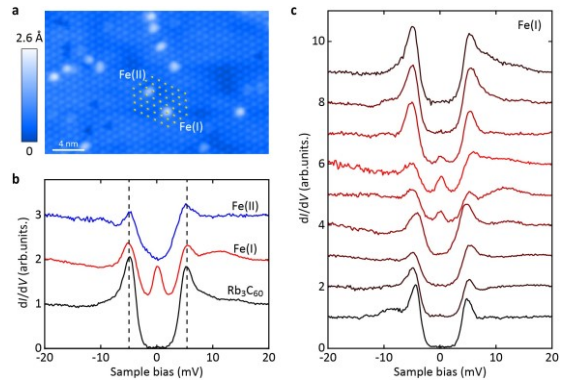

**Figure 4 Magnetic impurity (Fe)-induced bound states. a**, STM topography (28 nm × 18 nm, $V$ = 2.0 V, $I$ = 30 pA) of superconducting $Rb_3C_{60}$ films upon intentional post-deposition of tiny Fe adatoms. The yellow dots indicate the topmost $C_{60}$ molecules. **b**, Comparison of tunneling d$I$/d$V$ spectra averaged over single Fe(I), Fe(II) impurities and locations far from any impurities. Black vertical dashes denote the energy positions of the superconducting gap. **c**, Linecut d$I$/d$V$ spectra taken at equal separations (0.7 nm) across an isolated Fe(I) impurity. The red curve is measured justly on the impurity site. Set point: $V$ = 30 mV and $I$ = 200 pA.

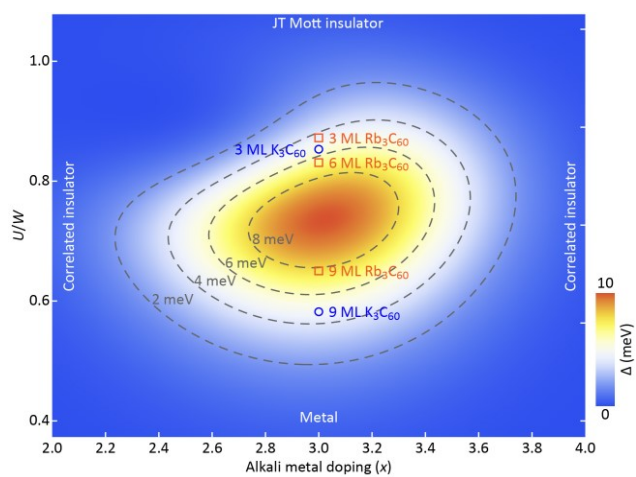

**Figure 5 Unified phase diagram of $\Delta$ variation with electronic correlations *U* and doping.** The empty circles and squares distinctively mark the experimental $\Delta$ measured in $K_3C_{60}$ and $Rb_3C_{60}$ films, respectively. Contour plots of $\Delta$ with a separation of 2 meV are shown in gray dashes. Note that the fulleride superconductivity is always peaked at half-filling at any specific *U/W*, with *W* denoting the $t_{1u}$ bandwidth.

*Supplementary Materials for*

**Merohedral disorder and impurity impacts on superconductivity of fullerenes**

Shu-Ze Wang[1*], Ming-Qiang Ren[1*], Sha Han[1], Xu-Cun Ma[1,2], Qi-Kun Xue[1,2,3,4], Can-Li Song[1,2*]

[1]*State Key Laboratory of Low-Dimensional Quantum Physics, Department of Physics, Tsinghua University, Beijing 100084, China*

[2]*Frontier Science Center for Quantum Information, Beijing 100084, China*

[3]*Beijing Academy of Quantum Information Sciences, Beijing 100193, China*

[4]*Southern University of Science and Technology, Shenzhen 518055, China*

*Both authors contributed equally to this work.

[†]*To whom correspondence should be addressed. Email: clsong07@mail.tsinghua.edu.cn*

**This PDF file includes:**

**Figures (S1-S6) and captions**

Fig. S1. Superconductivity in Cs and Rb co-doped fulleride $RbCs_2C_{60}$

Fig. S2. Coherence peak amplitude versus superconducting gap $\Delta$

Fig. S3. Magnetic vortices and pseudogap in 9 ML $Rb_3C_{60}$

Fig. S4. Insulating ground states in monolayer and bilayer $Rb_3C_{60}$

Fig. S5. Superconductivity and impurities in $K_xC_{60}$ away from half-filling

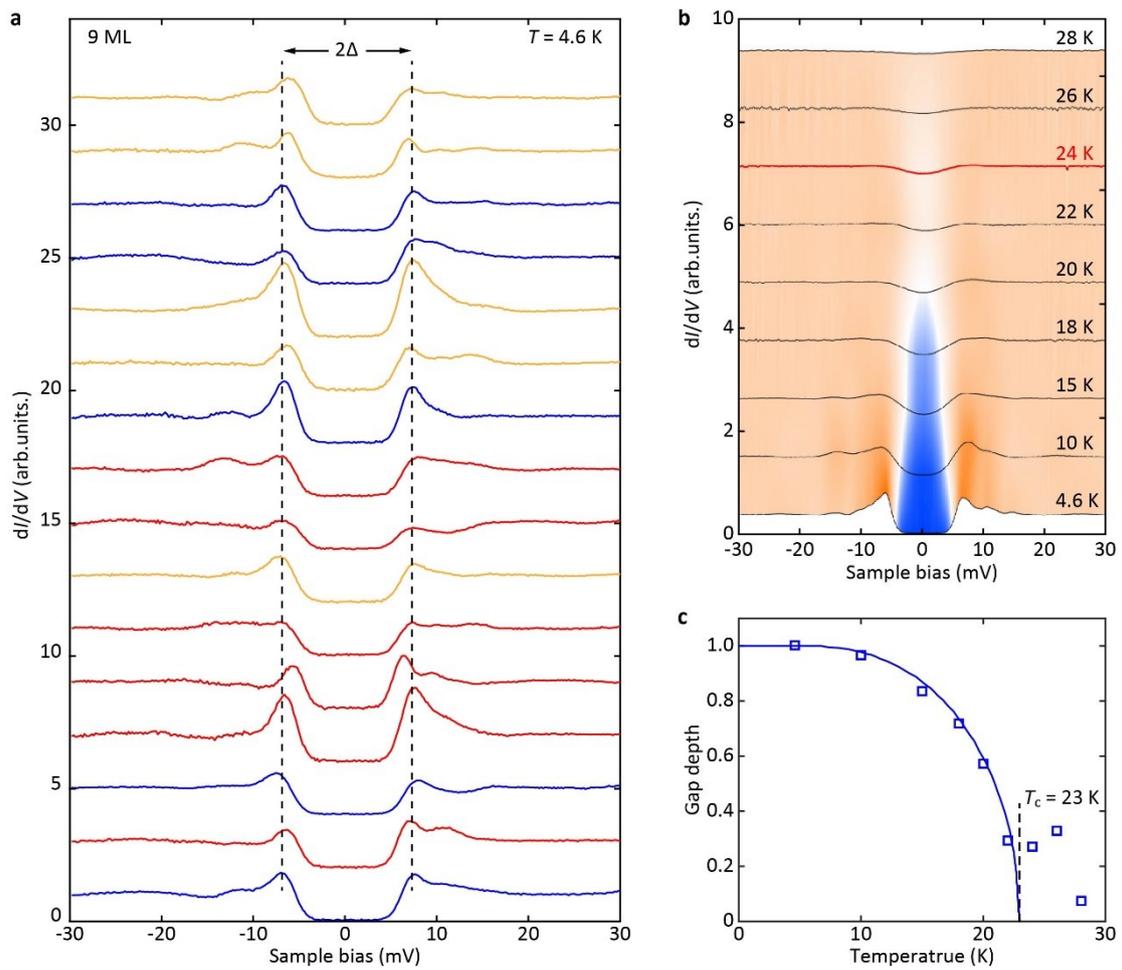

**Fig. S1. Superconductivity in Cs and Rb co-doped fulleride RbCs$_2$C$_{60}$. a,** Grid d$I$/d$V$ spectra (4 pixels × 4 pixels) in a field of 20 nm × 20 nm. Setpoint: $V$ = 30 mV and $I$ = 200 pA. Red, yellow and blue curves were measured in either of the two standard C$_{60}$ merohedral domains and regions between them. No discernible difference is revealed in the merohedrally disordered regions (blue curves). Black vertical dashes are guide to the eye. **b,c,** Temperature dependent tunneling d$I$/d$V$ spectra and gap depth of 9 ML RbCs$_2$C$_{60}$ film, presenting an apparent pseudogap (red curve) above $T_c$ = 23 K.

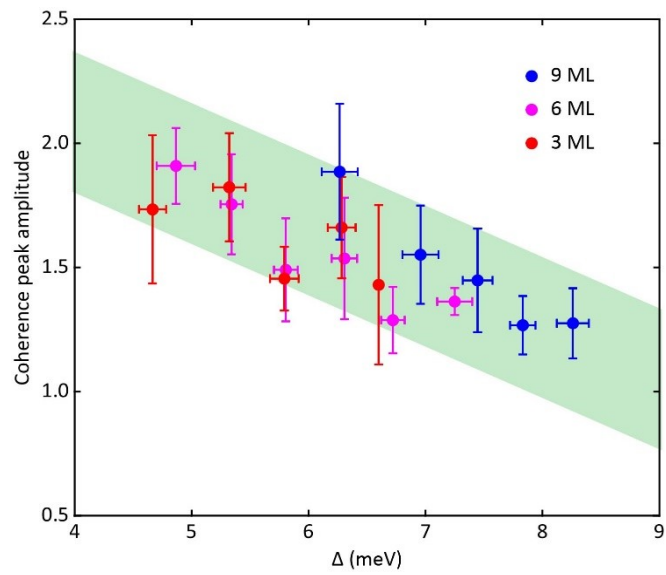

**Fig. S2. Coherence peak amplitude *versus* superconducting gap $\Delta$.** All data were measured at 4.6 K on superconducting $Rb_3C_{60}$ films at varied thicknesses/regions, binned and averaged for five ranges of gap magnitude $\Delta$. The statistical errors indicate the standard derivation of $\Delta$ and coherence peak amplitude obtained from different regions. The green parallelogram is a guide to the eye.

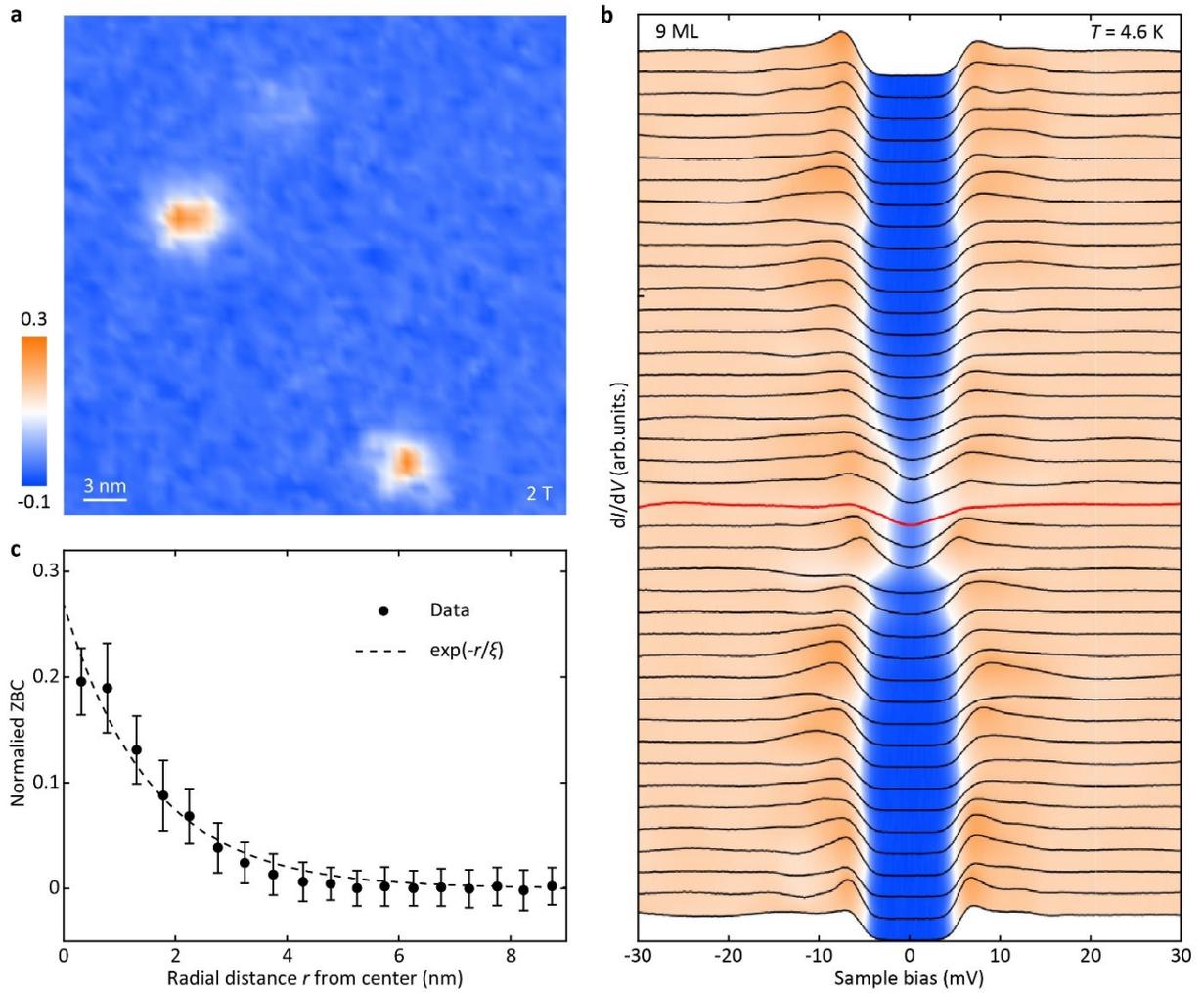

**Fig. S3. Magnetic vortices and pseudogap in 9 ML Rb$_3$C$_{60}$. a**, Normalized zero-bias conductance (ZBC) map (35 nm × 35 nm) showing two vortices at 2 T. In the vicinity of each vortex core (orange regions), the normalized ZBC is invariably less than unity and hints at the existence of pseudogap. **b**, A series of normalized d$I$/d$V$ spectra taken at equal separations along a 20-nm trajectory across one vortex. The red curve denotes the pseudogap within vortices. **c**, Radial dependence of normalized ZBC around the magnetic vortices. Dashed line shows the best fit of the data to an exponential function, yielding a superconducting coherence length of $\xi$ = 1.5 ± 0.2 nm in Rb$_3$C$_{60}$. The statistical errors of normalized ZBC indicate its standard derivation obtained from different redial angles.

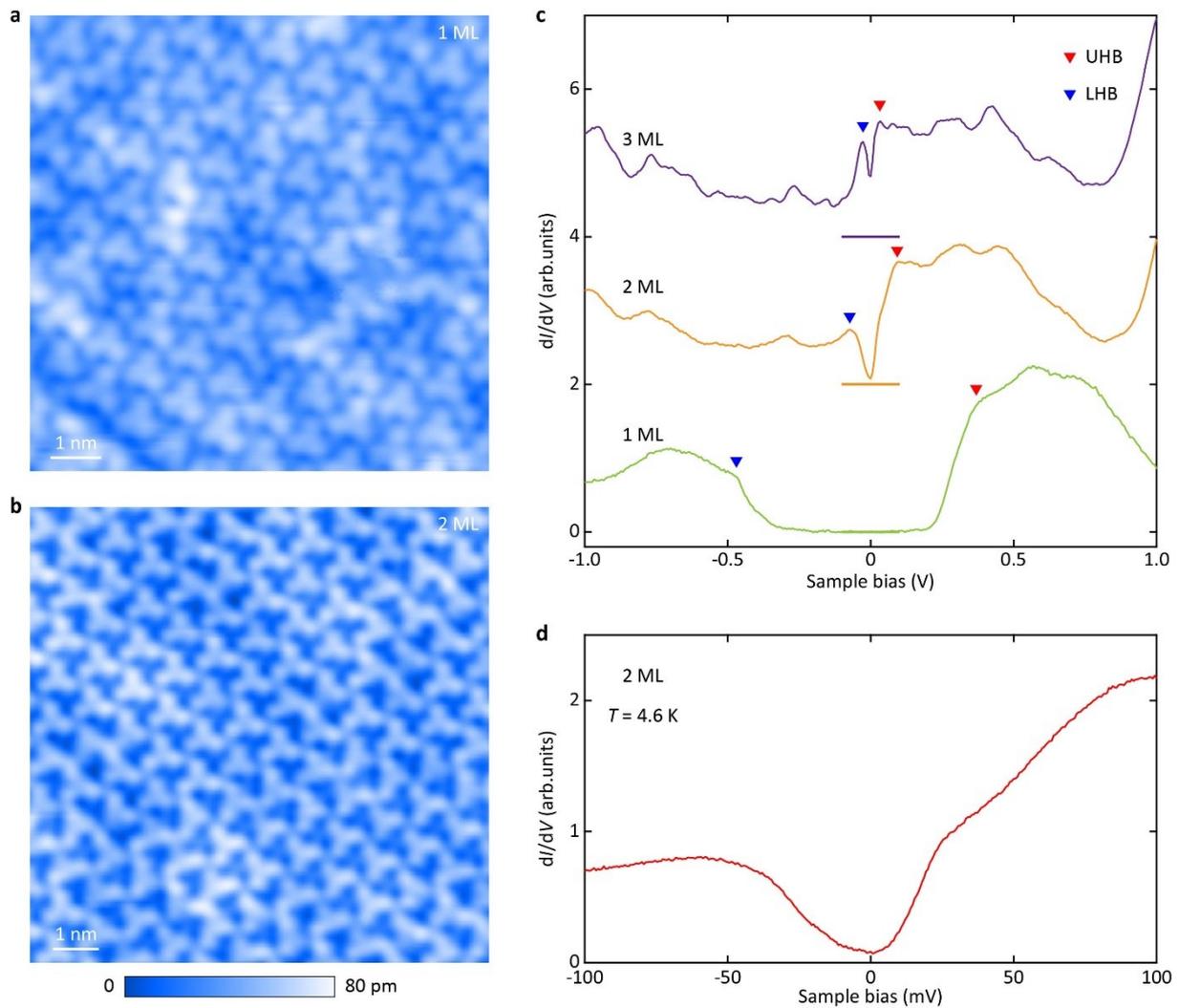

**Fig. S4. Insulating ground states in monolayer and bilayer Rb$_3$C$_{60}$. a**, STM topographies ($V$ = 1.0 V, $I$ = 20 pA) of monolayer (9 nm × 9 nm) and bilayer Rb$_3$C$_{60}$ films (10 nm × 10 nm), respectively. **c**, Typical d$I$/d$V$ spectra on monolayer, bilayer and trilayer Rb$_3$C$_{60}$ films. At the 2D limit, the enhanced electronic correlations result in a dip in electronic DOS and even insulating energy gap around $E_F$, characterized by the Hubbard $U$ between the upper Hubbard band (UHB, red triangles) and lower Hubbard band (LHB, blue triangles). Setpoint: $V$ = 1.0 V, $I$ = 100 pA. **d**, Lower-energy-scale d$I$/d$V$ spectrum exhibiting no evidence of superconductivity in bilayer Rb$_3$C$_{60}$ films. Setpoint: $V$ = 1.0 V, $I$ = 100 pA. In contrast to K$_3$C$_{60}$ counterparts (reference 33), no superstructure is found in Rb$_3$C$_{60}$ films at the two-dimensional (2D) limit. Nevertheless, monolayer and bilayer Rb$_3$C$_{60}$ are insulating due to the increased electronic correlations, analogous to the K$_3$C$_{60}$ counterparts.

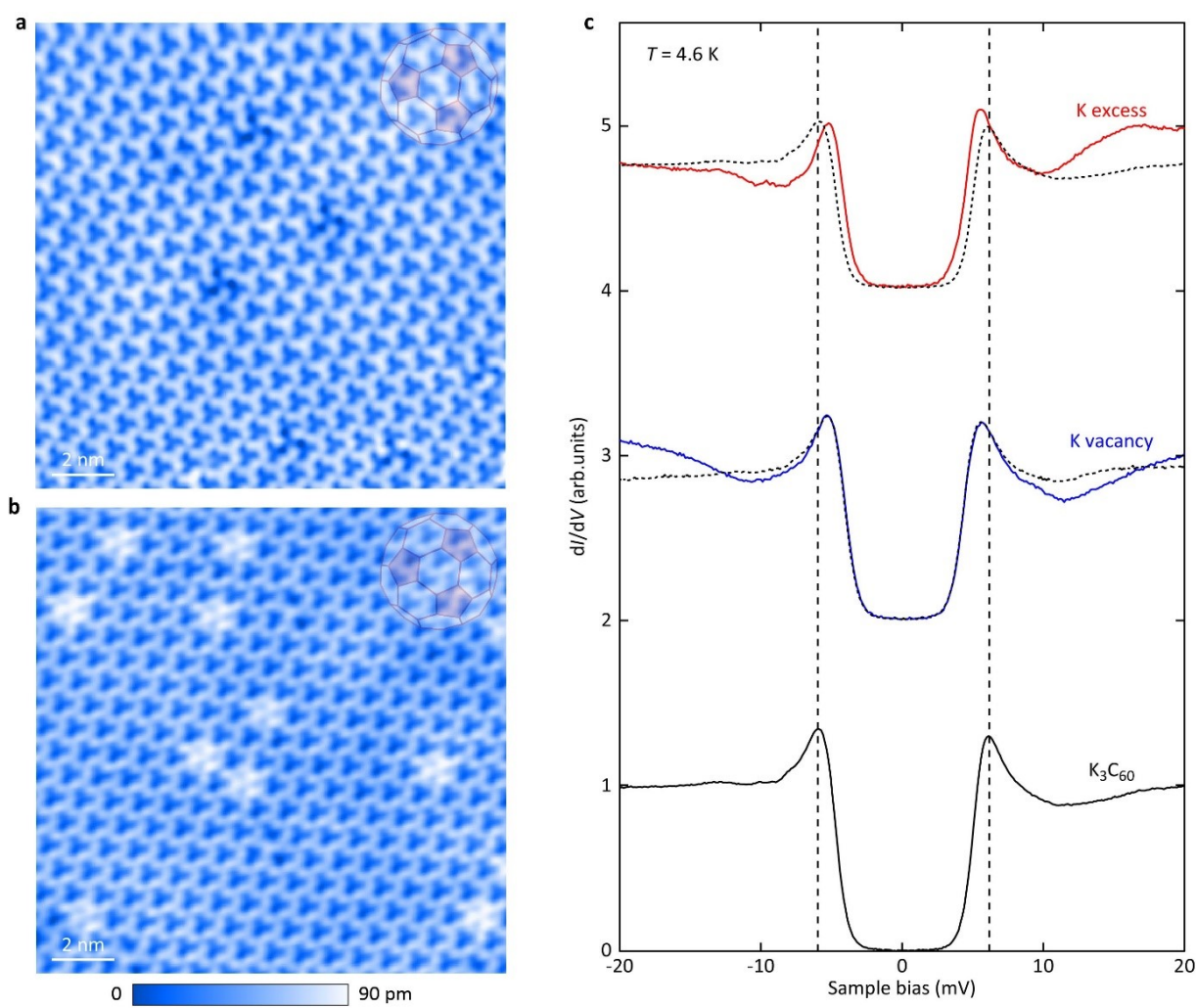

**Fig. S5. STM imaging and superconductivity of K-doped fullerenes away from half-filling. a,b**, STM topographies (15 nm × 15 nm, $V$ = 1.0 V, I = 10 pA) of $K_xC_{60}$ with tetrahedral K vacancies ($x < 3$) and octahedral excess K adatoms ($x > 3$), respectively. Insets illustrate the schematic view of one standard orientation of $C_{60}$ molecules along the [111] direction. Apparently, all molecules including those near K defects (i.e. K vacancies and excess adatoms) are of the same orientation, involving no merohedral disorder. **c**, Site-resolved d$I$/d$V$ spectra on non-magnetic K defects (blue and red curves), defect-free regions of $K_xC_{60}$ (dashed lines) and stoichiometric $K_3C_{60}$ (black line) at 4.6 K. For every impurity type, at least five impurities have been measured that consistently reveal no in-gap bound states. The vertical dashed lines are guides to the eye.